\documentclass[10pt,letterpaper]{article}

\usepackage{graphicx}

\usepackage{amsmath,amssymb}

\usepackage{subfig}

\usepackage{epstopdf}

\usepackage{bm}

\usepackage{subfig}

\usepackage[english]{babel}

\usepackage{authblk}

\usepackage{fullpage}

\usepackage[colorlinks=true,bookmarks=false,citecolor=blue,urlcolor=blue]{hyperref}

\newcommand{\wmp}{\hat{\bm{w}}}

\newcommand{\argmin}{\operatornamewithlimits{argmin}}

\newcommand{\argmax}{\operatornamewithlimits{argmax}}

\newcommand{\bw}{\bm{w}}

\newcommand{\DN}{\bm{D}_N}

\newcommand{\AN}{\bm{A}_N}

\newcommand{\xn}{\bm{u}_n}

\newcommand{\NA}{\mathrm{NA}}

\newcommand{\kmin}{k_{\mathrm{min}}}

\begin{document}


\title{Compressive and Adaptive Millimeter-wave SAR}

\author[1]{Alex Mrozack\thanks{alex.mrozack@duke.edu}}
\author[2]{Martin Heimbeck}
\author[1]{Daniel L. Marks}
\author[3]{Jonathan Richard}
\author[2,4]{Henry O. Everitt}
\author[1]{David J. Brady}
\affil[1]{Duke Imaging and Spectroscopy Program, Duke University ECE Dept, PO Box 90291, Durham, NC 27708, USA}
\affil[2]{Charles M. Bowden Research Center, Army Aviation \& Missile RD\&E Center, Redstone Arsenal, AL 35898, USA}
\affil[3]{IERUS Technologies, 2904 Westcorp Blvd, Suite 210, Huntsville, AL 35805, USA}
\affil[4]{Duke University Physics Dept, PO Box 90305, Durham, NC 27705, USA}
\maketitle
%
%
%
%
%
%
%




\begin{abstract}
We apply adaptive sensing techniques to the problem
of locating sparse metallic scatterers using high-resolution, frequency
modulated continuous wave W-band RADAR. Using a single detector, a
frequency stepped source, and a lateral translation stage, inverse synthetic
aperture RADAR reconstruction techniques are used to search for one
or two wire scatterers within a specified range, while an adaptive algorithm
determined successive sampling locations.  The two-dimensional location of each scatterer is thereby identified with sub-wavelength accuracy in as few as 1/4 the number of lateral steps required for a simple raster scan.  The implications of applying this approach to more complex scattering geometries are explored in light of the various assumptions made.
\end{abstract}




\section{Introduction}

One of the greatest challenges facing the millimeter wave (MMW) and especially the terahertz (THz) imaging communities is the restriction posed by the requirement to use expensive point detectors.  The impressive scans of obscured objects frequently reported in the MMW and THz literature are usually obtained through slow raster scanning of a source, object, or detector, often taking hours or days to complete\cite{TISAT,ISAM,sar_gpr}.  Although source power and detector sensitivity are improving, the rate-limiting factor remains the desired signal-to-noise ratio (SNR) of the scattered signal coupled with the limited mechanical scanning speed and/or the associated mechanical settling time before an acquisition can begin.  Although mechanical scanning is often the only practical strategy for obtaining an image of a complex scene with diverse spatial content, there are many problems where the imager is only being used to find isolated or sparsely-distributed scatterers in a visually opaque host.  For example, one might wish to find nails behind wallpaper or metallic plumbing behind sheetrock.  For such problems, it is impractical to raster scan large areas.

Coherent sources are common in the MMW and THz imaging bands, so synthetic aperture imaging may be used to overcome the limitation of sensing with a single, large, and expensive transceiver.  The synthesized aperture can either use a diverging beam, as is typically done in synthetic aperture radar\cite{sar_gpr}, or a quasi-optical system with a converging beam, as is done in optical coherence tomography (OCT) to increase penetration depth in scattering media\cite{TISAT,ISAM}.  The scanning time for synthetic aperture systems using classical processing techniques [e.g. \cite{TISAT,ISAM,sar_gpr}] could be reduced if more powerful sources and more sensitive detectors were used, but that would greatly increase system cost.  However, if the number of spatial samples could be reduced, it becomes more practical to use less expensive sources and detectors, especially if rapid mechanical scanning and efficient data processing are combined to reduce the time required to estimate a scene.

Compressive sensing (CS) consists of estimation of $P$ signal values from $N<P$ measurements. CS has been used to improve sampling efficiency and increase temporal resolution in many imaging systems.  A mathematical construct known as the Restricted Isometry Property (RIP) rigorously allows compressive measurement of high dimensional data on lower dimensional spaces\cite{candes_rip}. By this construct, CS terminology has become widely used in analysis of signals sparse under $l_1$ constraints\cite{Donoho2, candes1}.  Previous demonstrations of compressive planar terahertz imaging \cite{chan}, 3D holographic and millimeter wave tomography \cite{bradyCH, cull}, and scanned interferometric tomography \cite{compressive_oct} are particularly relevant to the $l_1$ construct.  Unfortunately, the sensing paradigms that provably obey RIP generally rely either on random sensing modes\cite{random_matrices}, sensing modes that obey strict rules of incoherence\cite{duarte_cs_review}, or special mode sets such as tight frames and related matrix constructs\cite{bajwa,sapiro}.

To make imaging as presented in \cite{TISAT} practical, we demonstrate here a compressive sampling technique to locate sparse point scatters with fewer spatial samples.  Compressive scanning for visible wavelength synthetic aperture imaging has already been presented in \cite{compressive_oct}, but two aspects of that work are undesirable for MMW synthetic aperture imaging of point targets.  One is the usual OCT assumption that the beam is in the confocal space, which makes imaging of point targets compressively difficult due to the tightly confined lateral distribution of the beam.  The other is the use of a random sampling set.  While this may be a good strategy for certain cases, a random set of modes is not a highly optimized sampling strategy.  An optimized sampling strategy would measure adaptively by taking into account the measurements already made to choose the next measurement from the available set of modes in an optimized way.

Here we report a method for scanning a synthetic aperture adaptively when the scene complexity (i.e. number of point scatterers) is assumed to be known. First the measurement model is presented.  The model is then taken as an input to the general framework specified by \cite{MacCay,Carin}.  The measurement model is then simplified to allow for easier implementation in the adaptive framework.  Experimental results are finally discussed, and the extension of this work to more complex targets is considered in light of the assumptions made.

\section{Adaptive Sensing}
\subsection{Method}\label{sec:method}
\begin{figure}
\centering
\includegraphics[width=4in]{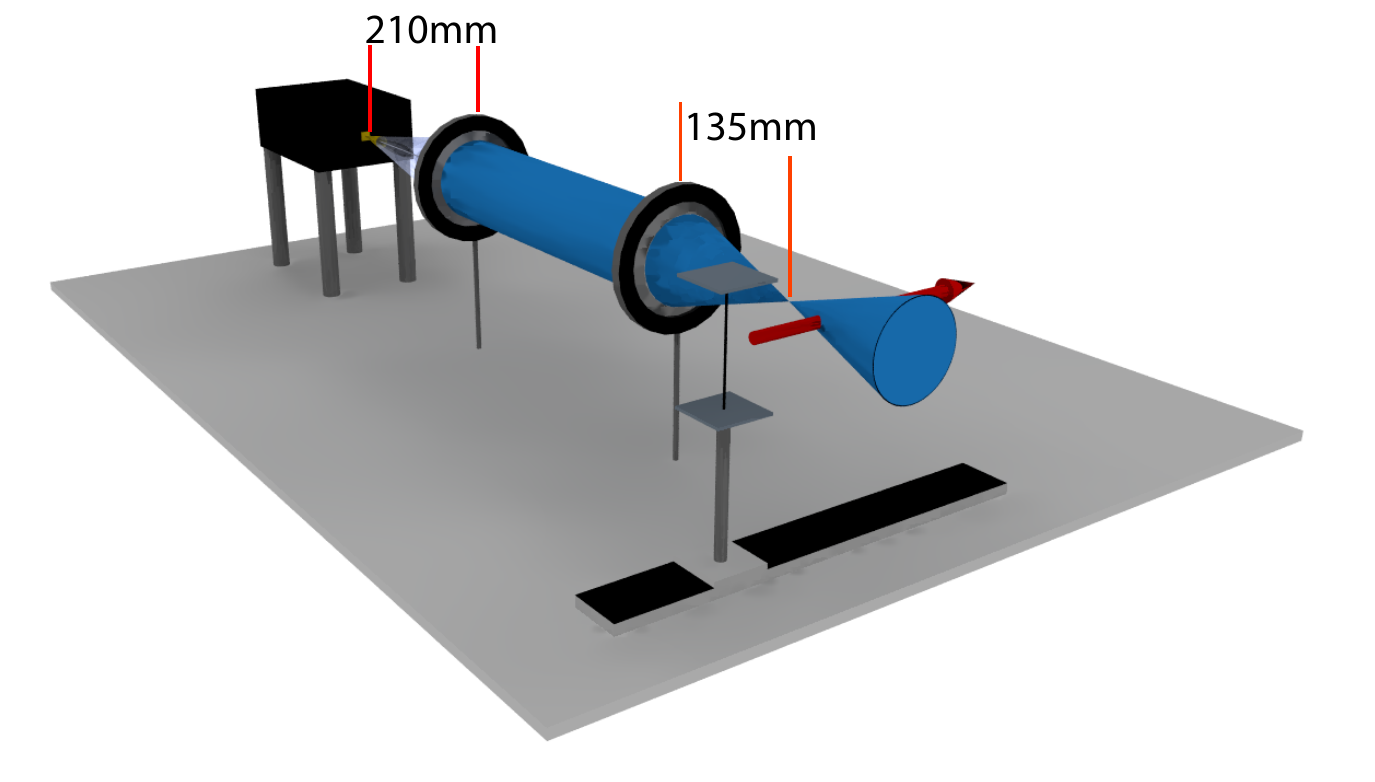}
\caption{The system diagram.  A transceiver generates a Gaussian beam transmit and receive gain pattern focused by two lenses in a 4f configuration.  A translation stage then moves the target through the beam to generate multiple lateral measurements.}

\label{fig:setup}

\end{figure}

Our goal is to locate a sparse array of scatterers as efficiently as possible using a synthetic aperture system with a single transceiver capable of sweeping frequency over a wide bandwidth attached to a linear stage.  The transceiver produces a beam which is focused via a quasi-optical system into a sample, creating a Gaussian beam.  For a synthetic aperture system of this nature, a measurement includes both input (or system) parameters $\xn$ such as system $\NA$, wavenumber, etc., and object parameters $\bm{w}$ such as scatterer position and scattering strength. The $n^{\mathrm{th}}$ measurement is then 
\begin{align}
g(\bm{u}_n)= M(\bm{u}_n;\bm{w})+\eta
\end{align}
where $M$ is the measurement forward model which maps the input and object parameters to the dataset, and $\eta$ is a signal independent and identically distributed (i.i.d.) white Gaussian noise source $\eta \sim \mathcal{N}(\bm{0},\beta^{-1}\bm{I})$.  The aforementioned Gaussian beam optics, and first Born approximation scattering theory, allow for $M(\bm{u}_n;\bm{w})$ to be specified by $M_0\left(x_n,k_n,NA;\{x_i',z_i',q_i\}_{i=1}^P\right)$ defined by
\begin{align}& M_0\left(x_n,k_n,NA;\{x_i',z_i',q_i\}_{i=1}^P\right) = \sum_{i=1}^P \left[\frac{W_0}{W}\exp\left[-\frac{\left(x_i'-x_n\right)^2 }{W^2}-jkz_i'-jk\frac{\left(x_i'-x_n\right)^2}{2R}+j\tan^{-1}\frac{z_i'}{z_R}\right]\right]^2 q_i\label{eq:original_obj}
\end{align}
where $x$ is the lateral dimension scanned mechanically, $z$ is the depth direction scanned by frequency sweep, $W_0$ is the beam waist in the $z=0$ plane, $W_0(k_n,\NA) = \frac{2}{k_n\theta}$, $\theta$ is the divergence angle of the beam, $\theta(\NA) = \sin^{-1}\mathit{\NA}$, $z_R$ is the Rayleigh range, $z_R(k_n,\NA) = \frac{k_n}{2W_0^2}$, $W$ is the beam waist for arbitrary $z$, $W(z_i') = W_0\sqrt{1+\left(\frac{z_i'}{z_R}\right)^2}$, and $R$ is the radius of curvature of the beam, $R(z_i') = z_i\left[1+\left(\frac{z_i'}{z_R}\right)^2\right]$\cite{ISAM}.  Therefore, the forward model depends on three system parameters - lateral location $x_n$ of the transceiver, wavenumber $k_n$ of the interrogating beam, numerical aperture $\NA$ of the interrogating optics - and three object parameters $\bw_i$ consisting of the complex scattering strength $q_i$ and the two dimensional location $(x_i',z_i')$ for each scatterer in the field\cite{ISAM}.  The square in the sum of Eq. \eqref{eq:original_obj} is from the fact that both the scene illumination and effective receiver gain pattern are assumed to have the same Gaussian beam mode.  The measurement system used in this study is shown in Fig. \ref{fig:setup}, and is discussed in detail in Sec. \ref{sec:setup}.  For the purposes of this section all that is necessary is knowledge of the analytical acquisition model $M$.

Given $M$, the goal is to measure adaptively until the scene ($\{x_i',z_i',q_i\}_{i=1}^P$) can be estimated reliably.  To do so we adopt the methodology of adaptive estimation similar to the adaptive classification procedures first specified by \cite{MacCay} and used for mine detection via electrostatics by \cite{Carin}.  The methodology relies on the specified measurement model and existing data to specify a current estimate and a next best measurement location.   

The collection of all $N$ measurements defines the dataset $\bm{D}_N = \{g(\xn)\}_{n=1}^N$ for which the probability density function is defined as
\begin{align}
p(\bm{D}_N|\bm{w}) \propto \exp\left[-\frac{\beta}{2}\sum_{n=1}^{N}\left| M\left(\bm{u}_n;\bm{w}\right)-g\left(\bm{u}_n\right)\right|^2\right].\label{eq:likelihood}
\end{align}

To obtain an estimate $\wmp$ for $\bm{w}$, a maximum a posteriori (MAP) estimate is used.  Given the posterior distribution
\begin{align}
p\left(\bm{w}|\bm{D}_N\right) \propto p\left(\bm{D}_N|\bm{w}\right)p_0 \left(\bm{w}\right),\label{eq:map}
\end{align}
the estimate must be
\begin{align}
\wmp = \argmin_{\tilde{\bm{w}}}\frac{\beta}{2}\sum_{n=1}^{N}\left|g(\bm{u}_n)-M(\bm{u}_n;\tilde{\bm{w}})\right|^2-\log p_0(\tilde{\bm{w}}),\label{eq:obj_function}
\end{align}
where $p_0(\bm{w})$ is a prior distribution on the object. In general $p_0(\bm{w})$ may be any appropriate distribution for the object.  For practical reasons to be explained shortly, the prior distribution is often chosen to be uniform or Gaussian with mean zero.  The experiments presented in this paper are concerned with detection of extremely sparse scatterers in two dimensions, lateral and depth, with the number of scatterers $P$ equal to 1 or 2.  When $P = 1$, the objective function is a measure of how well a single point with lateral position $x_1'$, depth position $z_1'$, and complex scattering strength $q_1 = {q_r}_1+\bm{i}{q_i}_1$ match the data given the model $M_0$, posing a 4-dimensional estimation problem.  For $P=2$ an 8 dimensional estimation problem must be solved, and for arbitrary $P$ \hspace{.5pt} $4P$ parameters must be estimated.  Now given a current estimate, the adaptive method needs to decide where to measure next given the current estimate and $\DN$.  To achieve this goal, two primary assumptions will be made.  The first is that the posterior distribution defined by Eq. \eqref{eq:map} is approximately Gaussian about mean $\wmp$ with inverse covariance matrix $\AN$
\begin{align}
p\left(\bw|\DN\right) \propto \exp\left[\frac{1}{2}(\bw-\wmp)^H \AN(\bw-\wmp)\right]\label{eq:posterior}.
\end{align}  
This is essential because maximizing the determinant of this precision matrix $\AN$ has information theoretic significance \cite{Carin}.  The exact significance of the determinant of the precision matrix is beyond the scope of this paper, but in simple terms it can be thought of as the aggregate precision over all of the parameters to be estimated.  Using the Gaussian assumption, the form for the precision matrix is trivially derived from Eq. \eqref{eq:posterior} as
\begin{align}
\log p\left(\bw|\DN\right) = -\frac{1}{2}\left(\bw-\wmp\right)^{H}\AN\left(\bw-\wmp\right)+C\\
\AN = -\nabla_{\bw}\nabla_{\bw}\log p\left(\bw|\DN\right),\label{eq:prec_eq_1}
\end{align}
where $C$ is an arbitrary constant.  From Eq. \eqref{eq:map}, Eq. \eqref{eq:prec_eq_1} becomes 
\begin{align}
\AN &= -\nabla_{\bw}\nabla_{\bw}\left(\log\left(p\left(\bm{D}_N|\bm{w}\right)\right)+\log\left(p_0 \left(\bm{w}\right)\right)\right).\label{eq:prec_eq_2}
\end{align}

Since Eq. \eqref{eq:likelihood} describes $p\left(\bm{D}_N|\bm{w}\right)$, it could be substituted directly into Eq. \eqref{eq:prec_eq_2} yielding a final form for the precision matrix.  However, for this application it is necessary for the matrix to depend only on the current estimate of the object and the locations of the measurements previously taken.  This can be achieved by the second necessary assumption that the model is linear about the estimate.  Taylor expanding $M$ about $\wmp$
\begin{align}
M(\xn;\bw)\approx M(\xn;\wmp)+(\bw-\wmp)^H \nabla_{\bw} M(\xn;\bw)|_{\wmp}\\
\bm{v}\left(\xn;\wmp\right) = \nabla_{\bw} M\left(\xn;\bw\right)|_{\wmp},
\end{align}
where $\bm{v}\left(\xn;\wmp\right)$ is the slope of the model about the estimate $\wmp$ at location $\xn$.  Of course this is a bad assumption for general models, but as the estimate is refined the assumption becomes better since changes in the estimate should be getting smaller if the algorithm is to converge.  Eq. \eqref{eq:likelihood} is then applied to Eq. \eqref{eq:prec_eq_2} using the new form for the model, and the precision matrix becomes
\begin{align}
\AN \approx \beta\sum_{n=1}^{N} \bm{v}(\xn;\wmp)\bm{v}(\xn;\wmp)^H-\nabla_{\bw}\nabla_{\bw}\log p_0 (\bw)\label{eq:prec_eq_3}
\end{align}
which depends only on the current estimate $\wmp$ and previous measurement locations $\{\xn\}_{n=1}^N$.  These properties, and the assumption that $\wmp$ does not change drastically from measurement to measurement, allow for the precision matrix from the next measurement to be approximated as
\begin{align}
\bm{A}_{N+1} = \AN+\beta\bm{v}(\bm{u}_{N+1};\wmp)\bm{v}(\bm{u}_{N+1};\wmp)^H
\end{align}
before the next measurement is taken, assuming $\nabla_{\bw}\nabla_{\bw}\log p_0 (\bw)$ is a constant or zero as happens when $p_0(\bw)$ is uniform or Gaussian as specified earlier.  An optimal choice of $\bm{u}_{N+1}$ would maximize the determinant of the precision matrix $\bm{A}_{N+1}$ \cite{Carin} which is the solution to 
\begin{align}
\bm{u}_{N+1} = \argmax_{\hat{\bm{u}}_{N+1}}\bm{v}(\bm{u}_{N+1};\wmp)^H\bm{A}_{N}^{-1}\bm{v}(\bm{u}_{N+1};\wmp).
\end{align}
The procedure is iterated until convergence in the change of the determinant of the precision matrix is achieved.
\subsection{Model Approximations}
\begin{figure} 
	\hspace{-.7in}\begin{tabular}{ccc}
\includegraphics[width=2.5in]{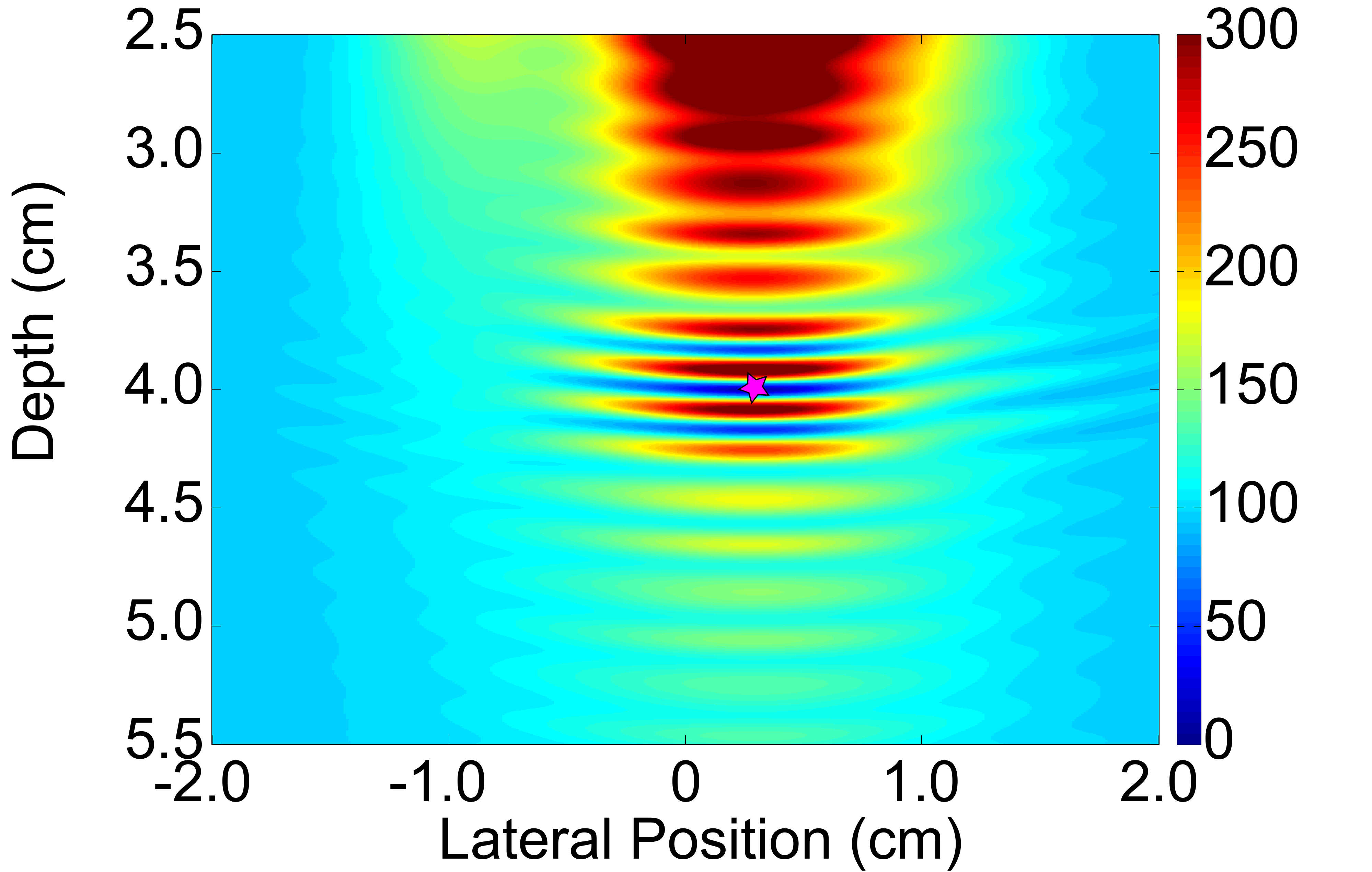} & \includegraphics[width=2.5in]{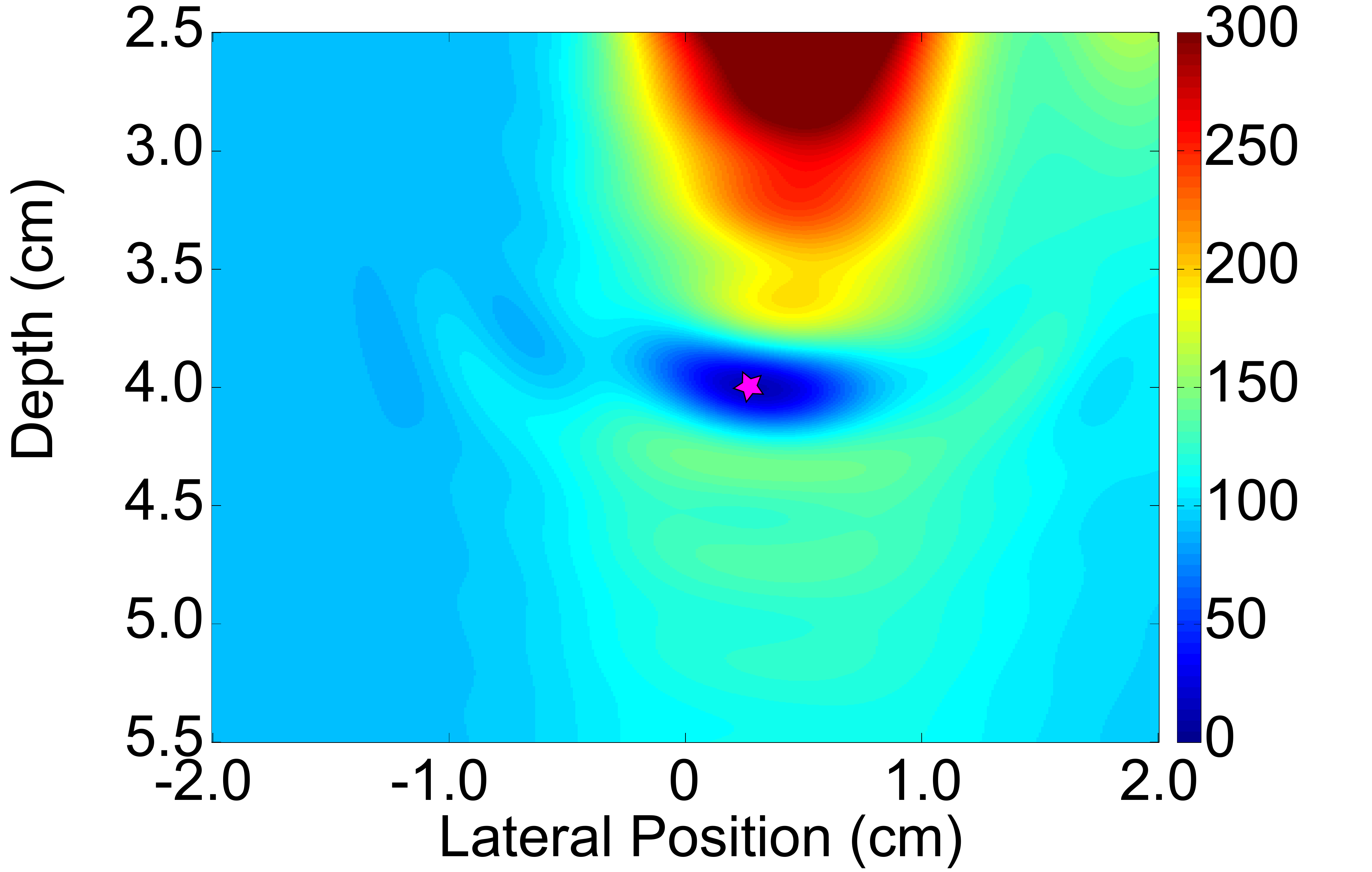} & \includegraphics[width=2.5in]{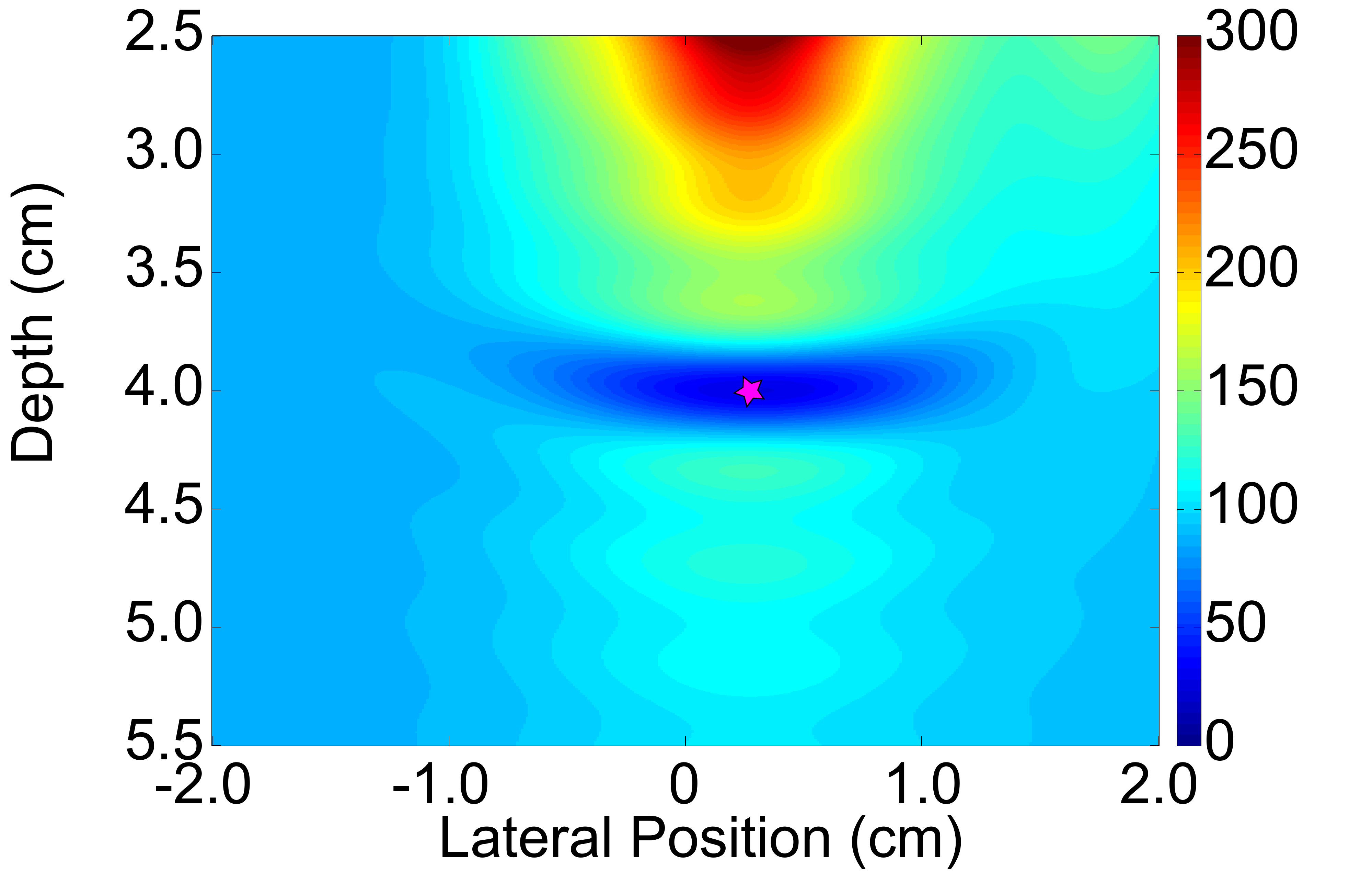}\\
(a) & (b) & (c)\\
\includegraphics[width=2.5in]{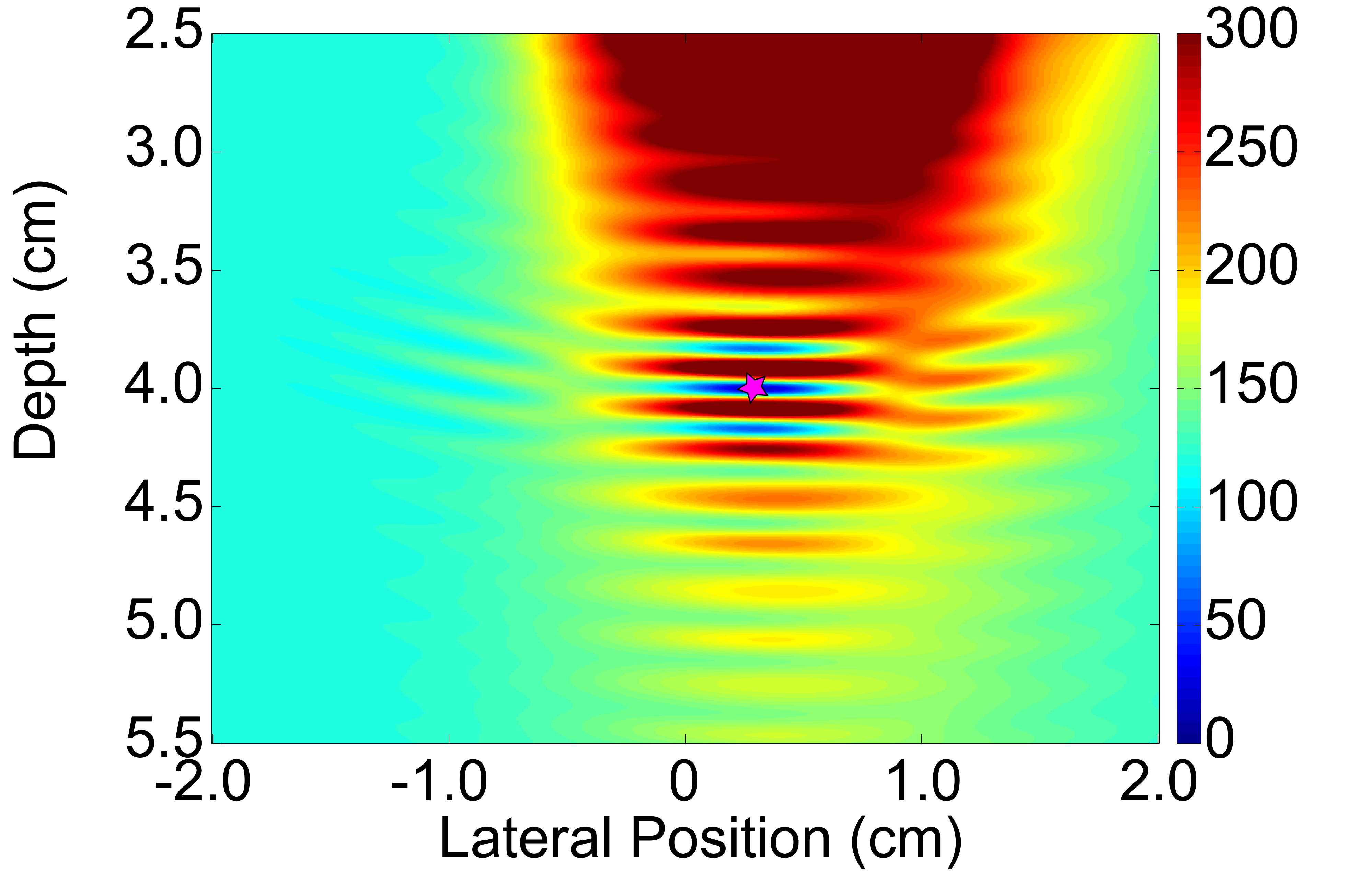} & \includegraphics[width=2.5in]{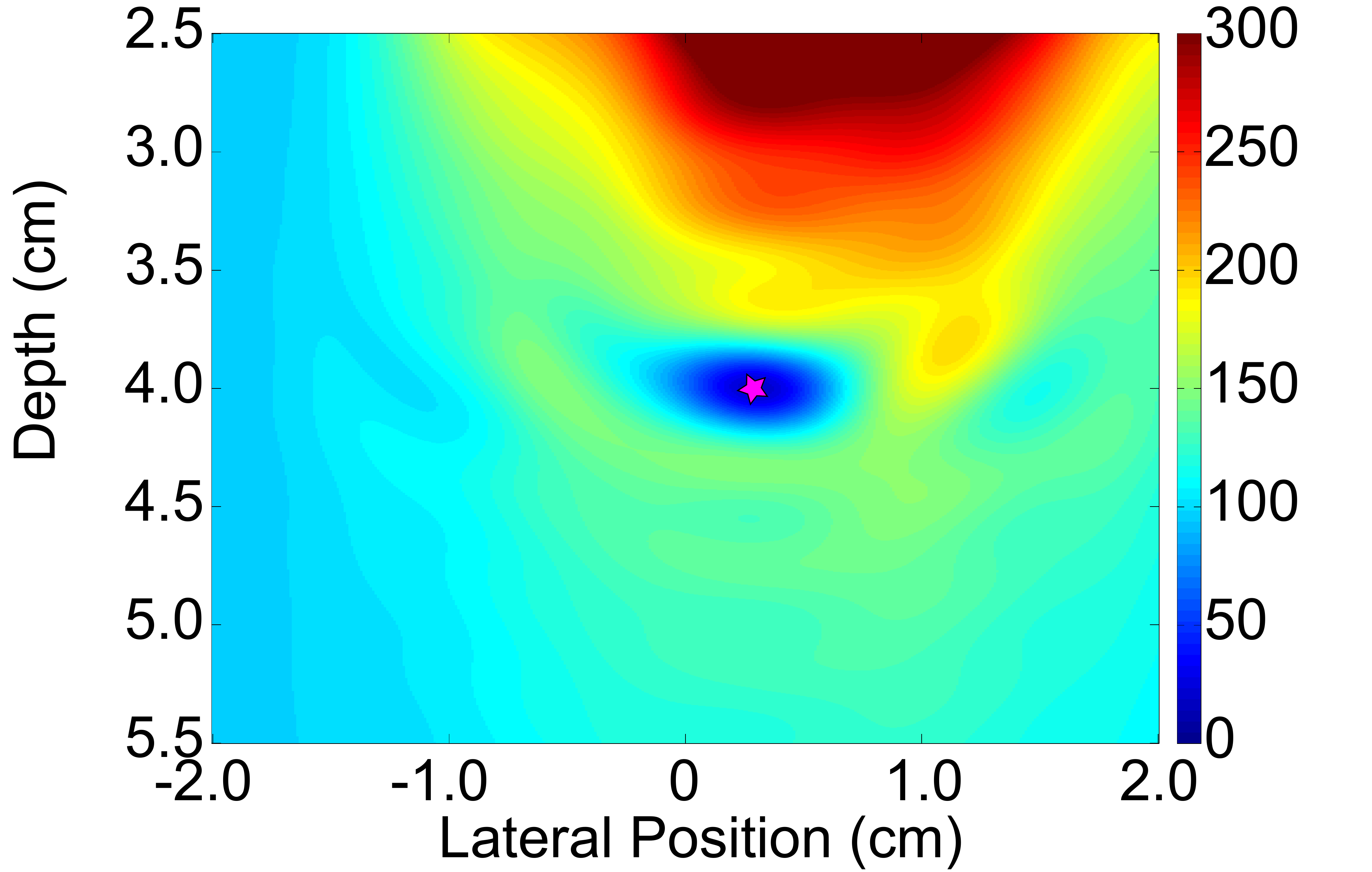} & \includegraphics[width=2.5in]{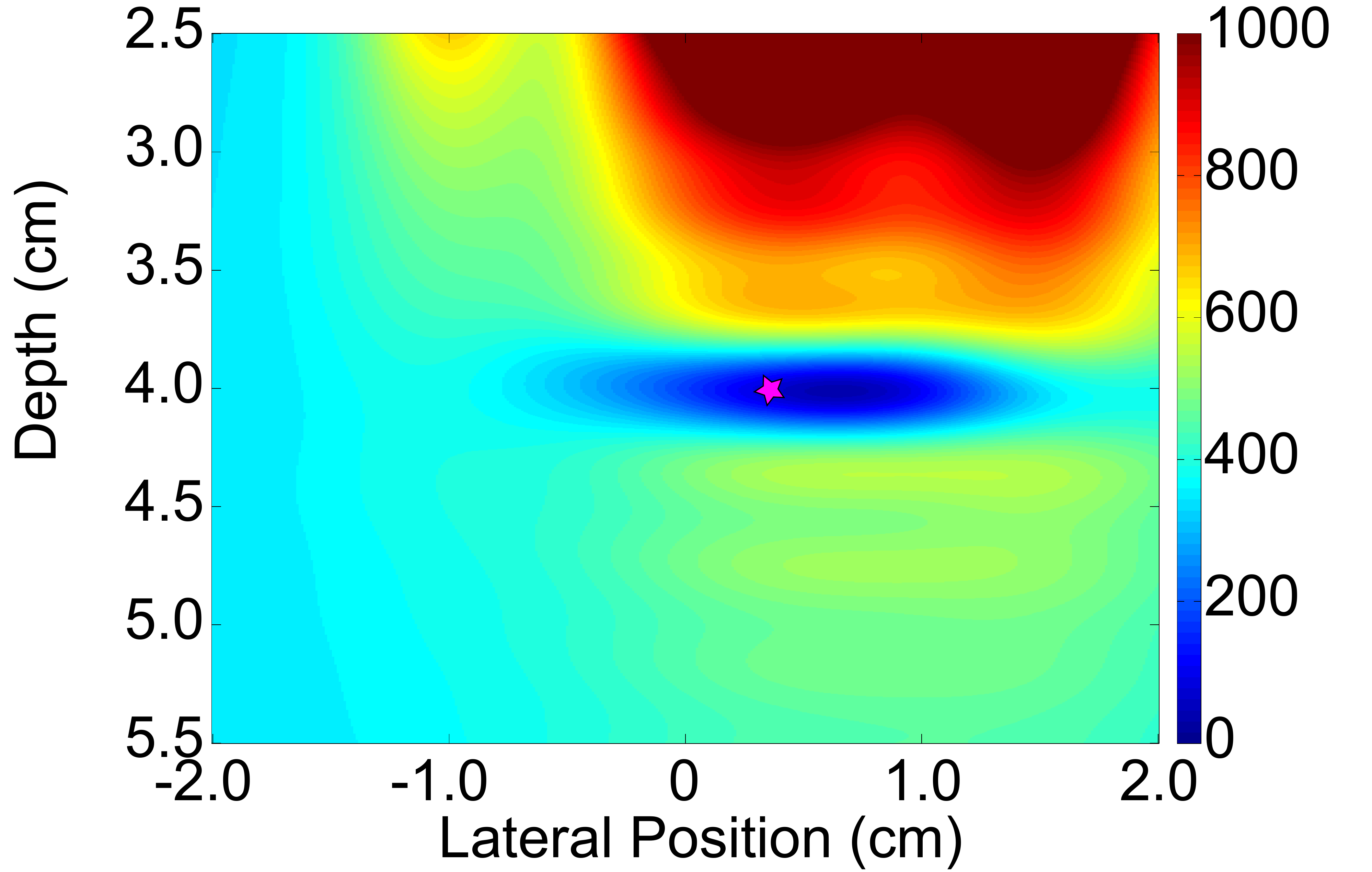}\\
(d) & (e) & (f)
	\end{tabular}
	\caption{Cross sections of the three objective functions in Eq. \eqref{eq:obj_function} with the model replaced with the model in Eq. \eqref{eq:original_obj} (plots (a) \& (d)), the depth demodulated model in Eq. \eqref{eq:no_ikz_obj} (plots (b) \& (e)), and the fully-demodulated model in Eq. \eqref{eq:no_ikx_obj} (plots (c) \& (f)).  The true location of the scatterer is indicated by the star. The top row is for a single point scatterer in which the scattering density is held constant and the two position variables of the objective function are varied.  The bottom row searches for the first of two point scatterers holding constant the scattering density of the first scatterer and all parameters of the second scatterer.  The true location of the sought scatterer is $x_1 =3$ mm and $z_i =4$ cm for all scenarios.  Note the reduced oscillations in (b) and (c) relative to (a), and in (e) and (f) relative to (d), as well as the inaccurate lateral location of the minimum in (f).}
	\label{fig:objective_funs}
\end{figure}

The specified adaptation procedure will only be successful if an optimization routine can find a reliable solution to Eq. \ref{eq:obj_function}.  Fig. \ref{fig:objective_funs}(a) shows a cut through the objective function specified in Eq. \eqref{eq:obj_function} after 8 simulated adaptive measurements, with $M$ specified by $M_0$ in Eq. \eqref{eq:original_obj} using a W-band system with a bandwidth of $75-110$GHz and $\NA$ of $.28$.  The cut is the height of the objective function for varying estimates of $x_1'$ and $z_1'$ while holding the estimate of ${q_r}_1$ and ${q_i}_1$ constant.  The true location of the scatterer is $x_1'=3$ mm and $z_1'=4$ cm.  Ideally this function would be a smooth bowl, allowing for a simple optimizer to use the gradient to find the minimum at the true location.  Unfortunately, this is not the case for the objective function with $M$ represented by $M_0$ as the measurement model oscillates strongly because the measurement is made far from baseband ($i.e.$ $k \gg 0$).  

This can be corrected if we make the common OCT assumption that the scatterers are nondispersive, allowing us to demodulate the model and perform data matching against a smoother function.  Fig. \ref{fig:objective_funs}(b) shows cuts through the objective function for the same scenario as for Fig. \ref{fig:objective_funs}(a) with $M$ being specified not by $M_0$, but by
\begin{align}
\widehat{M}\left(x_n,k_n,NA;\{x_i',z_i',q_i\}_{i=1}^P\right) = \sum_{i=1}^P \left[\frac{W_0}{W}\exp\left[-\frac{\left(x_i'-x_n\right)^2 }{W^2}-j(k-\kmin)z_i'-jk\frac{\left(x_i'-x_n\right)^2}{2R}+j\tan^{-1}\frac{z_i'}{z_R}\right]\right]^2 q_i\label{eq:no_ikz_obj}
\end{align}
which demodulates the axial plane-wave component of the Gaussian beam by subtracting $\kmin z$ from the phase, where $\kmin$ is the wavenumber corresponding to the lowest frequency sampled.  Clearly this has greatly reduced axial oscillations in the objective function.

However, there is another strongly contributing phase term in the Gaussian beam equation: the transverse quadratic term in $x$.  OCT usually works in the confocal region of the beam and does not have this term.  To demodulate both the axial plane-wave and lateral quadratic terms of the Gaussian beam, $\kmin z + \kmin x^2$ are subtracted from the phase.  Fig. \ref{fig:objective_funs}(c) shows cuts through the objective function for the same scenario as for Fig. \ref{fig:objective_funs}(a) and (b) with $M$ being specified not by $M_0$, but by 
\begin{align}
\widetilde{M}\left(x_n,k_n,NA;x_i',z_i',q_i\right) = \sum_{i=1}^P\left[\frac{W_0}{W}\exp\left[-\frac{\left(x_i'-x_n\right)^2 }{W^2}-j(k-\kmin)z_i'-j(k-\kmin)\frac{\left(x_i'-x_n\right)^2}{2R}+j\tan^{-1}\frac{z_i'}{z_R}\right]\right]^2 q_i.\label{eq:no_ikx_obj}
\end{align}
This has further reduced the oscillations in the objective function and produces the bowl-like shape desired for rapid convergence of an optimization routine.

Now that we have an objective function for the $P=1$ case, we need to verify that our assumptions will hold for two scatterers, especially when they are close to each other and the interference between them is strong.  Consider a second scatterer at $x_2' = 7$ mm and $z_2' = 4$ cm, approximately one beam waist away from the first scatterer still at $x_1' = 3$ mm and $z_1' = 4$ cm.  Fig. \ref{fig:objective_funs}(d) shows a cut through the objective function with $M_0$ as $M$ in Eq. \eqref{eq:obj_function} after 12 simulated adaptive measurements.  The cut plots the height of the objective function for varying $x_1'$ and $z_1'$ while holding ${q_r}_1$, ${q_i}_1$, $x_2'$, $z_2'$, ${q_r}_2$, and ${q_i}_2$ constant.  As in the $P=1$ case, the objective function rapidly oscillates.  Although substituting $\widehat{M}$ for $M$ reduces these oscillations as before (Fig. \ref{fig:objective_funs}(e)), substituting $\widetilde{M}$ for $M$ (Fig. \ref{fig:objective_funs}(f)) produces an erroneous result:  the minimum of the function shifts from $3$ mm to $5$ mm.  While this shift is less than a beam waist, lateral resolution is the most important criterion for imaging at these relatively large wavelengths.  Since $\tilde{M}$ provides the smoothest objective and accurately locates the single scatterer, $M$ is represented by $\tilde{M}$ for the $P=1$ case.  However, since the scatterers cannot be reliably located in simulation for the $P=2$ case if $M$ is represented by $\tilde{M}$, $M$ is instead represented by $\hat{M}$.

\section{Experiment}
\subsection{Setup}\label{sec:setup}
The W-band transceiver shown in Fig. \ref{fig:setup} consists of a 6x frequency multiplication chain that upconverts a $12.50-18.33$ GHz frequency sweep from an Agilent N5222A vector network analyzer (VNA) to a $75-110$ GHz output. A small portion of the output signal is down-converted with a reference mixer and coupled into the VNA’s reference input for phase sensitive measurements. The detector portion consists of another mixer which down-converts the received W-band signal for the VNA’s measurement input. This monostatic source-reference-detector setup is realized with packaged frequency extenders from Virginia Diodes VNAX TXRX WR10.

A focused source was generated by a two lens confocal imaging system. The first lens collimated the 20 degree diverging beam from the W-band source, and the second lens focused the beam in the vicinity of the object depth. Note that the scatterer is placed near but not in the focal plane of this confocal configuration, and the ideal location depends on considerations of the siganl-to-scatter ratio (i.e. the ratio of returned signal power to background signal power, analogous to signal-to-clutter ratio in radar).  The scatterer is more detectable if placed closer to the focal plane, but it is in the detector's field-of-view for fewer lateral locations and requires more measurements to locate.  Conversely, placing the scatterer farther from the focal plane reduces the received signal but permits it to be located in fewer measurements.  Clearly the $\NA$ of the interrogating beam plays a critical role in system considerations:  higher $\NA$ offers greater resolution throughout the imaging volume and higher signal intensity at the focus, but the beam diverges more quickly and the signal-to-scatter ratio becomes low for a shorter depth-of-field.  

For our experiments, operating characteristics similar to those in \cite{TISAT} were chosen.  Point scatterers in 2D space were either 1 or 2 wires suspended behind the focus.  These scatterers were placed several centimeters beyond the focus of a $.28 \NA$ beam, providing a cm-scale beam waist and a cm-scale depth-of-field while still retaining enough detectable signal.  The transmitter/optical system/receiver were held fixed, and the wire scatterers were laterally scanned by a Newport translation stage with 1 mm resolution. At each location, the data collected consisted of the scattered signal received in response to a complete $75-110$ GHz frequency sweep.

After setting up the system, three calibration tasks were performed.  First, the VNA output was calibrated so that reflection from the horn impedance mismatch at the output of the frequency extender did not create large reflections and reduce the dynamic range of the detector. All subsequent scans, calibration or otherwise were taken with the calibrated VNA. Second, the data from a single, full, calibration scan was rotated such that the focus of the beam was in the zero phase plane of the data.  Third, the amplitude of the calibration data was altered to match the spectrum assumed by the model, and the gain correction. The amount of rotation and gain correction were saved for later application to adaptively acquired data.  The correct rotation for the data was determined by performing the ISAM algorithm on the full scan for various rotations until the tightest focus was achieved \cite{TISAT}.  The amplitude correction was performed by empirically ascertaining a function that matched the amplitude of synthetically generated data for the same assumed system parameters.  In this case, the amplitude correction applied was the following gain correction:
\begin{align}
g'(x_n,k_n,\NA;x_1',z_1',q_1') = g(x_n,k_n,\NA;x_1',z_1',q_1')G(k_n)\\
G(k_n) = \exp\left[\frac{k^2_{\mathrm{max}}}{k_n^2}\right]\label{eq:gain_correction}.
\end{align}

While the correction was chosen empirically, it was shown to improve location resolution for cases and reasons to be discussed next.  It is important to note here, however, that the gain correction did not and should not alter the phase of the measurements.  If phase corrections were required, then either the previous two calibration steps were done incorrectly or the Gaussian beam assumption was wrong.  Furthermore, while these calibration steps required a full scan, they remained stable throughout the majority of the acquisitions taken over the course of several days.  Therefore, the number of measurements stated below for a reliable estimate of the target location do not include the calibration measurements.



\subsection{Simulated Scans from Real Data: 1 and 2 Points}\label{sec:simulations}
\begin{figure}
\hspace{.5cm}\includegraphics[width=6in]{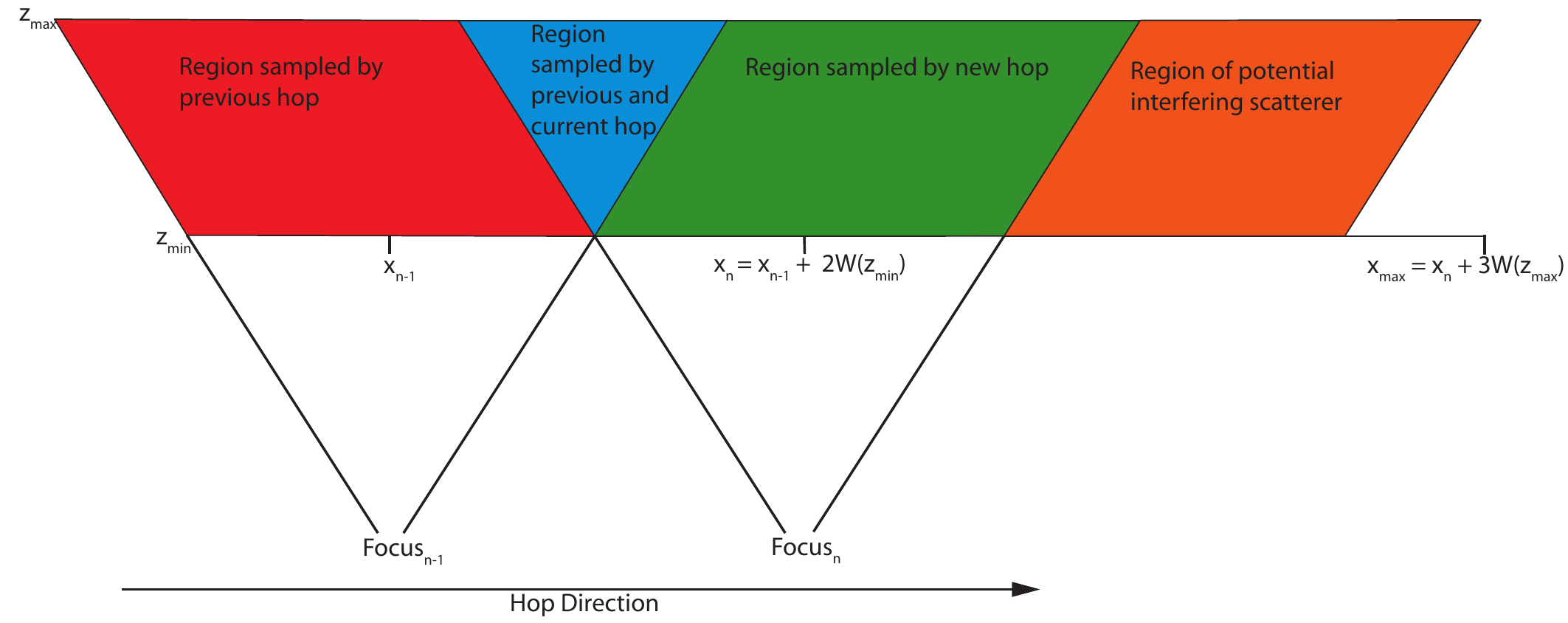}
\caption{The optimal hopping geometry showing the previously sampled region (red), the oversampled region (blue), the newly interrogated region of interest (green), and the region of possible interference for $P=2$(orange).  Operating farther from the focus yields more efficient sampling, but lower SNR.}
\label{fig:optimal_geometry}
\end{figure}

To ascertain the feasibility of the adaptive formalism, we first synthetically performed adaptive measurements on previously acquired complete datasets to ascertain how few lateral measurements were required to estimate the scatterer's position and with what accuracy.  The first dataset was of a single scatterer located at $x_1'=0$mm and $z_1'=4.8$cm, and two scenarios were simulated.  The first scenario adaptively scanned to estimate the location of a scatterer somewhere within a wide window ($-2cm<x_1'<2cm$ and $3cm<z_1'<7cm$).  The second scenario used an optimal hopping approach where the scanner would take large jumps until a single point scatterer entered the field-of-view.  This optimal hopping strategy could only locate a point scatterer within the parallelogram-shaped region illustrated in Fig. \ref{fig:optimal_geometry}.  For both scenarios, adaptation selected the next best lateral location to measure, and the frequency sweep at that new location will be added to the data previously collected to estimate the location and scattering strength of the target. (Only the lateral dimension was sampled adaptively because frequency sweeping is much faster than the time required to move the scanner.)  

\begin{figure}
\centering
\begin{tabular}{cc}
\includegraphics[width=2.2in]{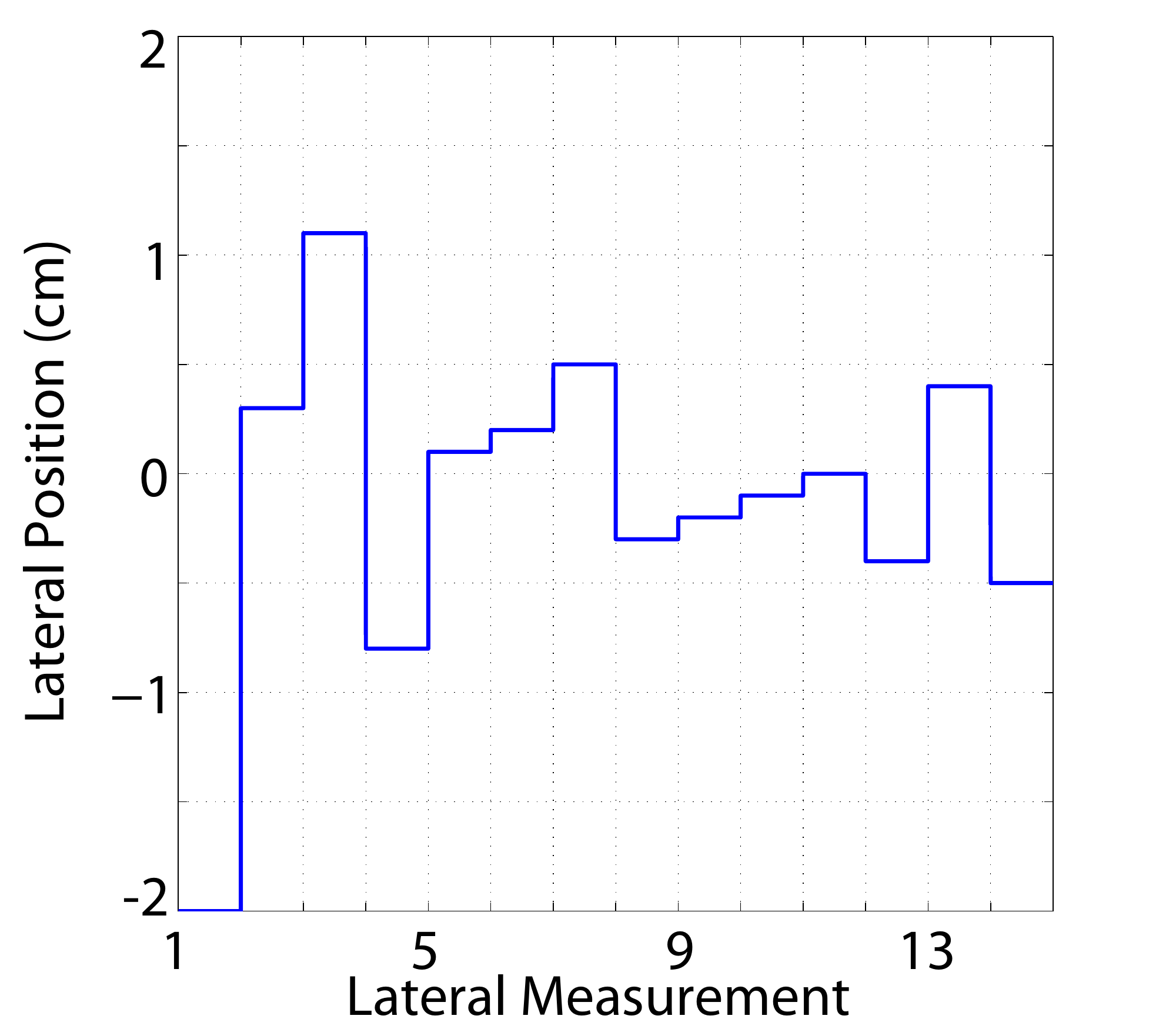} & \includegraphics[width=2.2in]{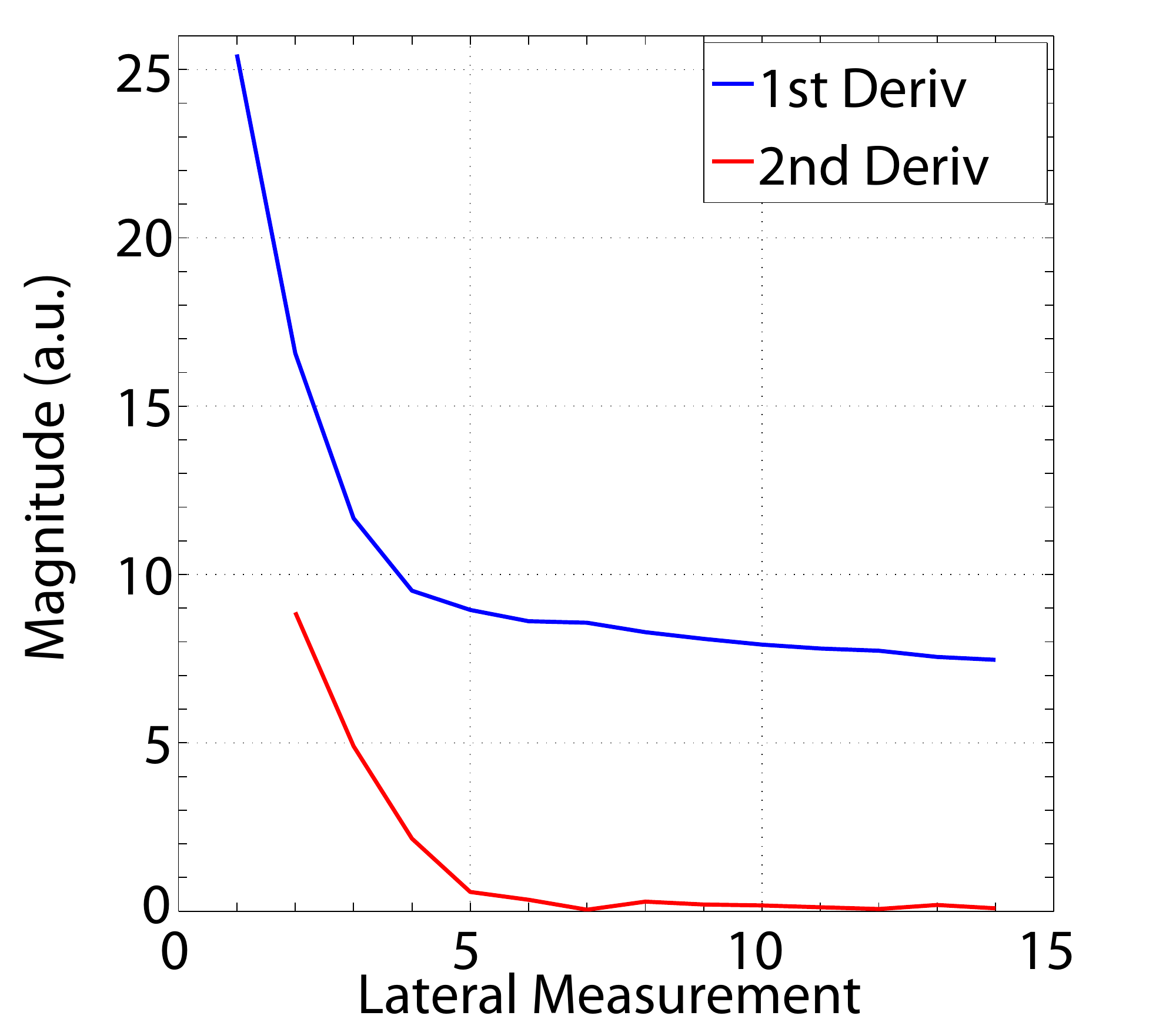}\\
(a) & (b)\\
\includegraphics[width=2.2in]{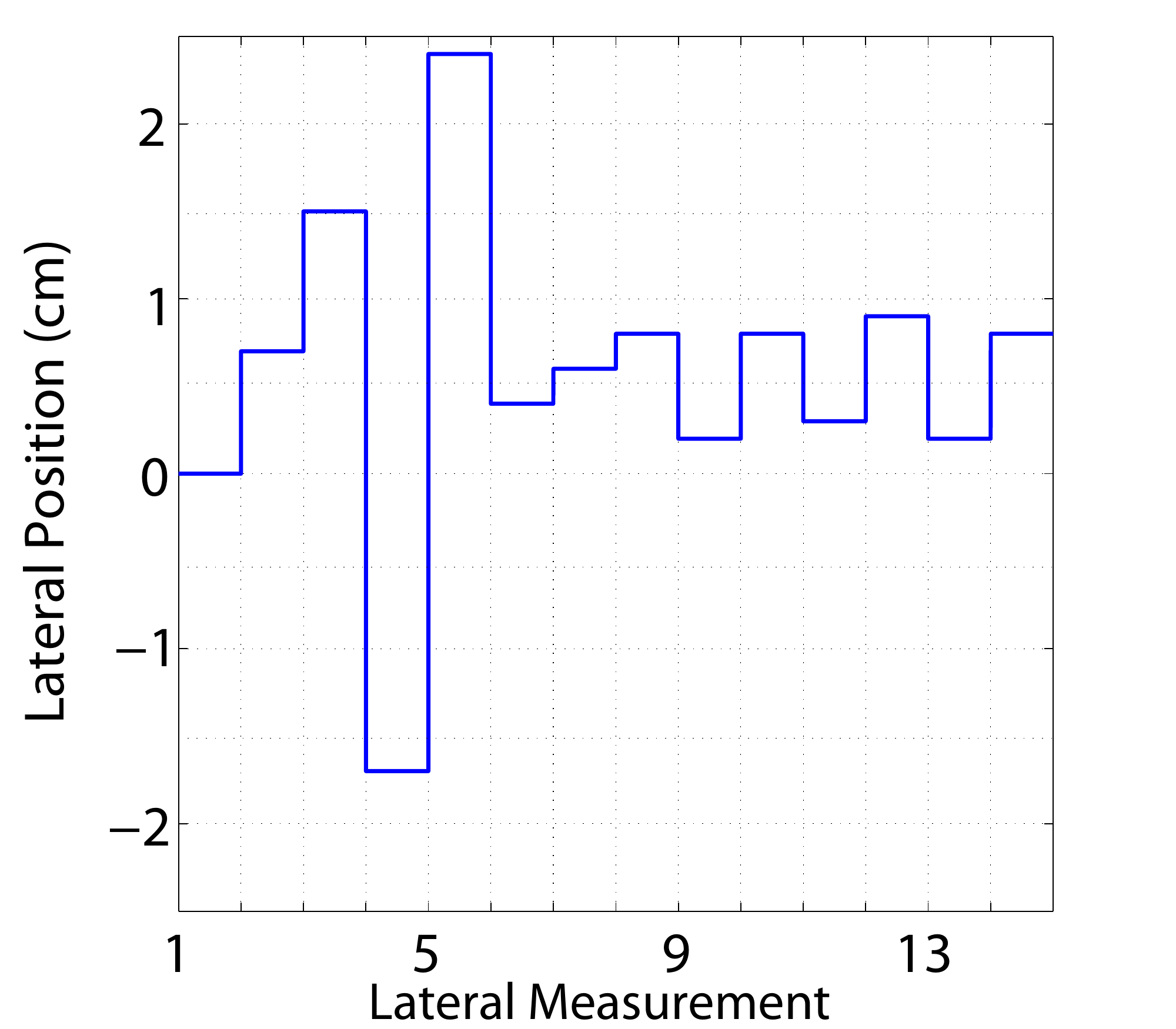} & \includegraphics[width=2.2in]{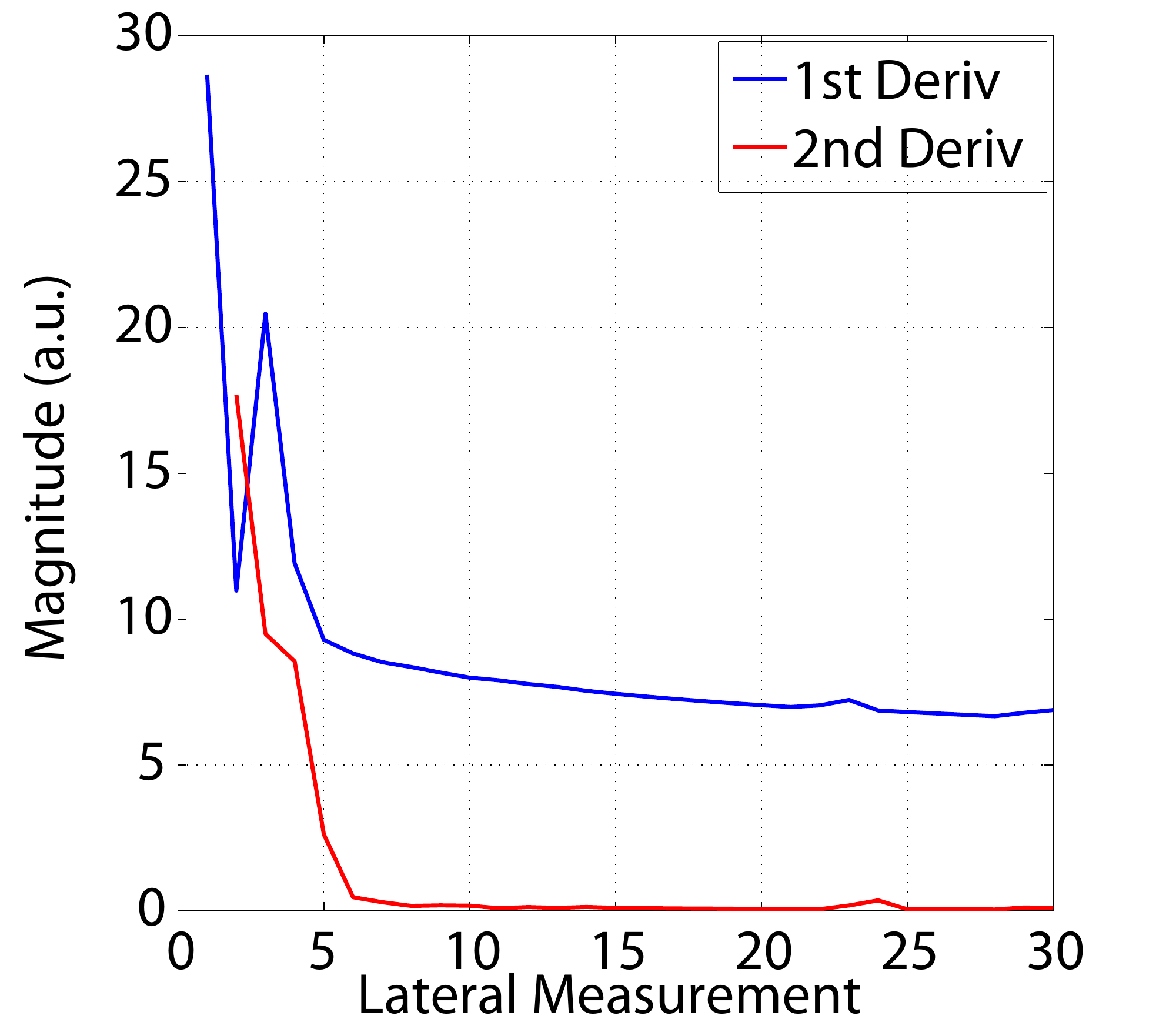}\\
(c) & (d)
\end{tabular}
\caption{A typical measurement path for a simulated (a) and fully adaptive acquisition (c), and the magnitude of the first and second derivative of the precision matrix versus measurement for that path (b) and (d), respectively.}
\label{fig:typical_run}
\end{figure}

Fig. \ref{fig:typical_run}(a) shows a typical adaptive path taken through the data for the wide window case; the optimal hop case behaves similarly but is more tightly constrained.  The simulated scanner started (lateral measurement 1) by jumping in the negative direction away from the starting location by a large amount.  Constrained not to repeat a measurement, the scanner then reversed direction and jumped a larger distance in the positive direction (lateral measurement 2), placing the object within the beam.  Next the scanner moved a smaller amount in the positive direction (lateral measurement 3) and kept the object in the beam.  However, upon recognizing the received signal was weaker, the scanner moved in the negative direction back to a location between the first and second position (lateral measurement 4).  Subsequent measurements placed the scanner closer to the lateral location of the object (lateral measurements 5-11) but contributed less information, and at some point the algorithm could have stopped collecting data altogether.  To see when convergence was reached and the algorithm could have stopped, consider the precision matrix $\AN$ and the derivatives of the $\AN$ with respect to lateral measurement shown in Fig. \ref{fig:typical_run}(b).  For the first few lateral measurements, the change in the determinant of $\AN$ was large, but with each successive measurement, the change in det($\AN$) became smaller.  After only $7$ measurements the second derivative of det($\AN$) vs. iteration was approximately zero, and little more information could be obtained about the scene.  At this point, a practical implementation of the algorithm would have stopped acquiring data.  To see what would happen if it did not stop, Fig. \ref{fig:typical_run} indicates that eventually there are no more locations to measure close to the scatterer, so the scanner begins to take larger steps oscillating around the scatterer (lateral measurements 12-15).  
%

\begin{figure}
\centering
\begin{tabular}{cc}
\includegraphics[width=2.2in]{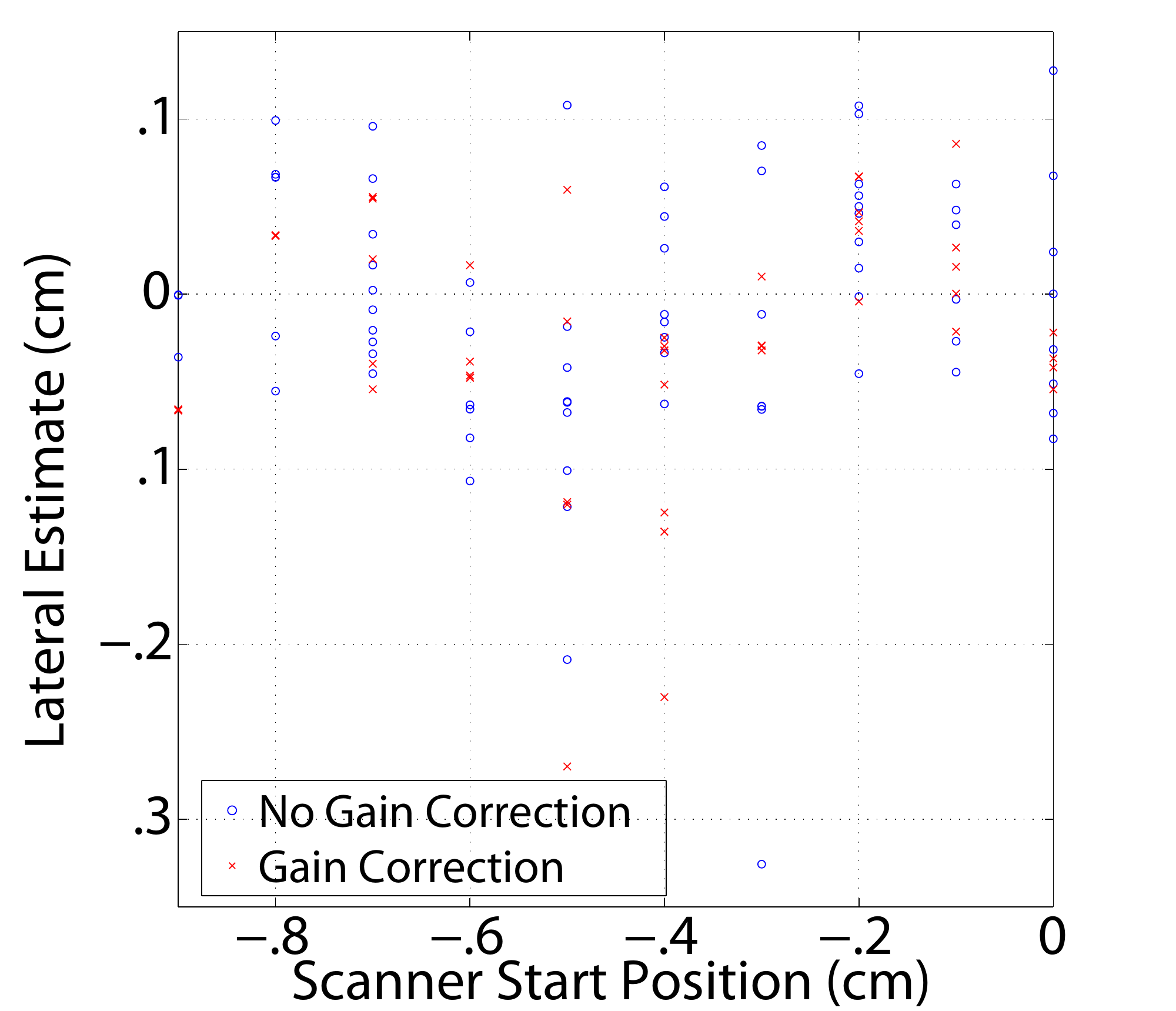} & \includegraphics[width=2.2in]{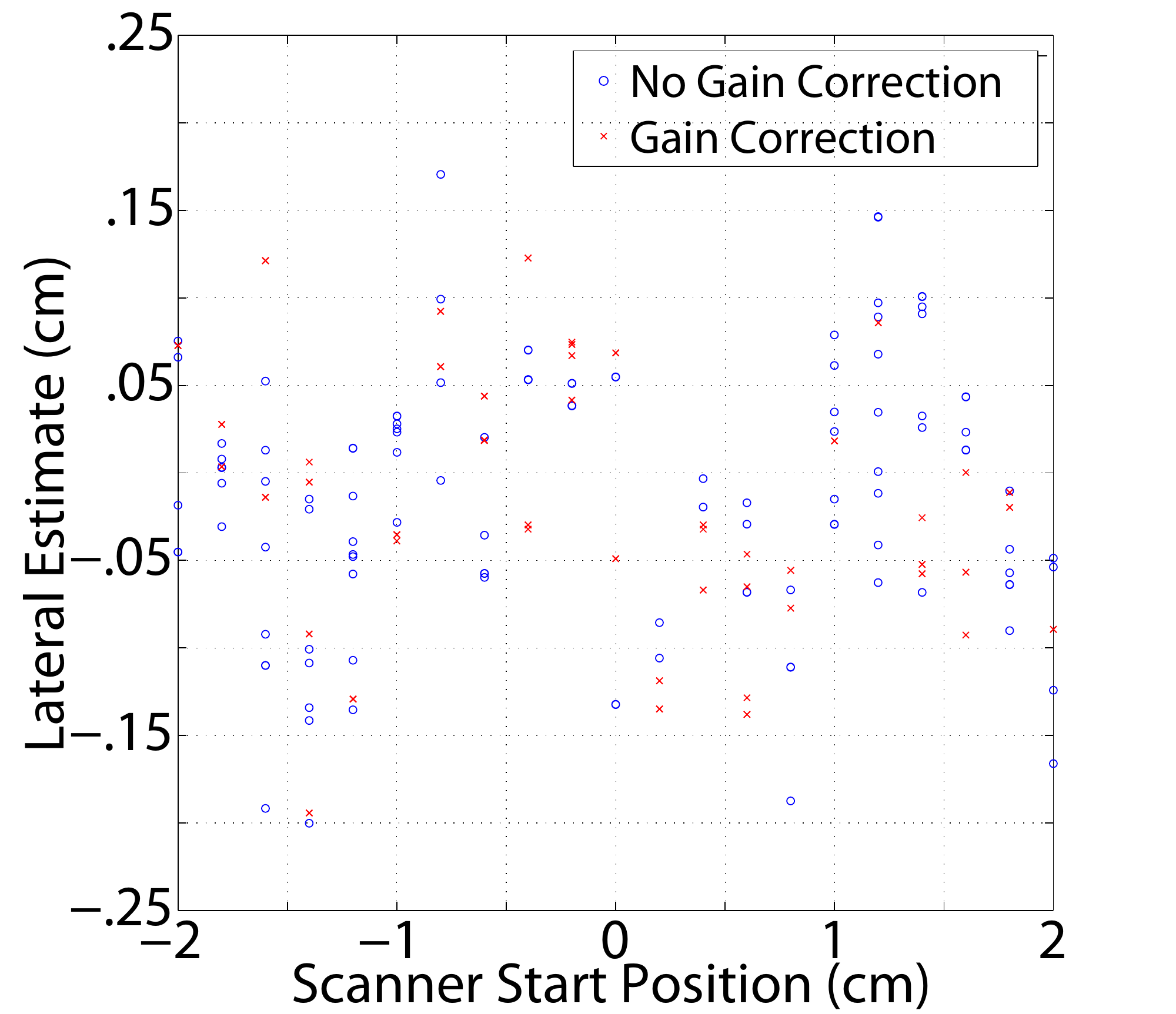}\\
(a) & (b)\\
\includegraphics[width=2.2in]{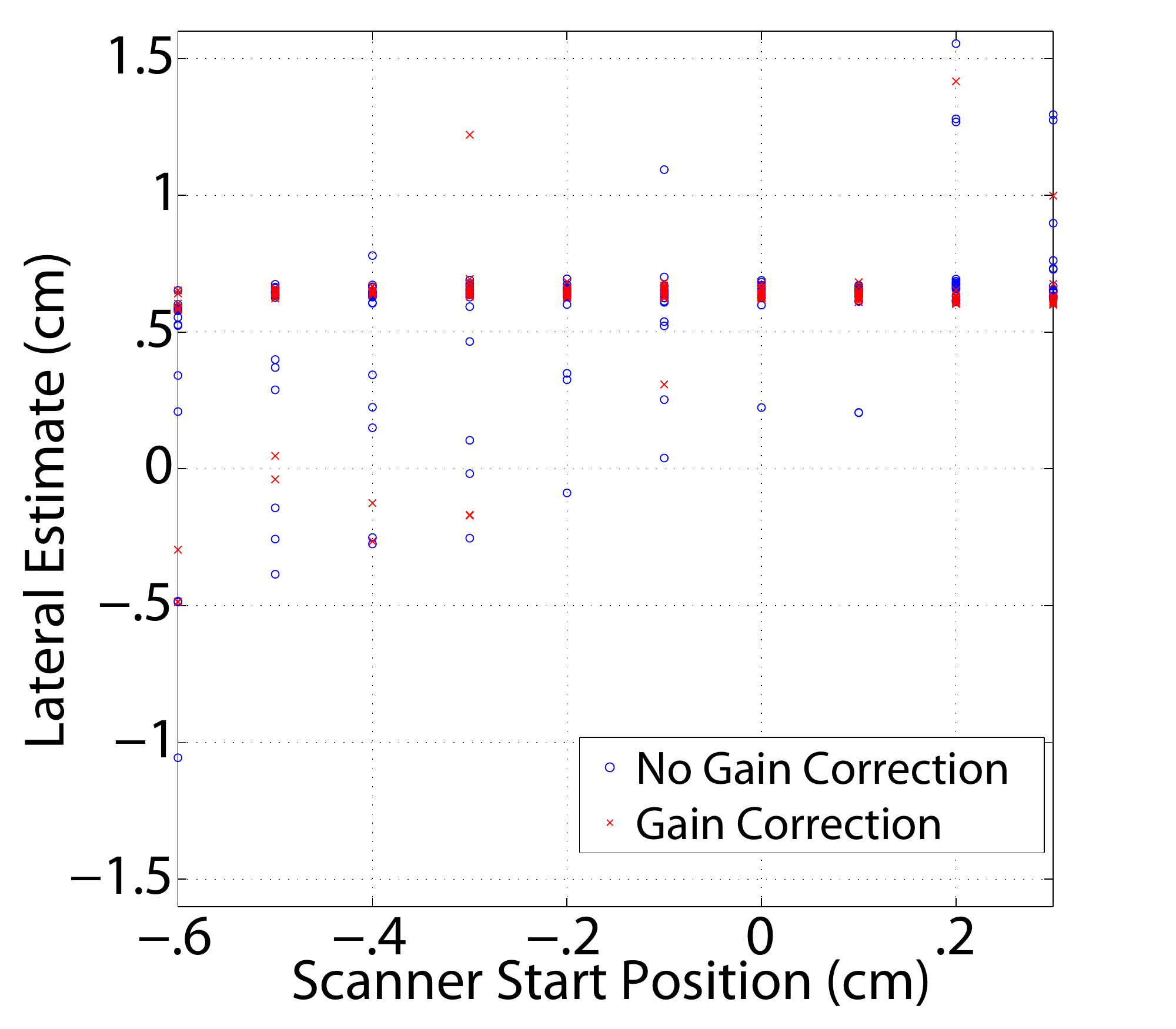} & \includegraphics[width=2.2in]{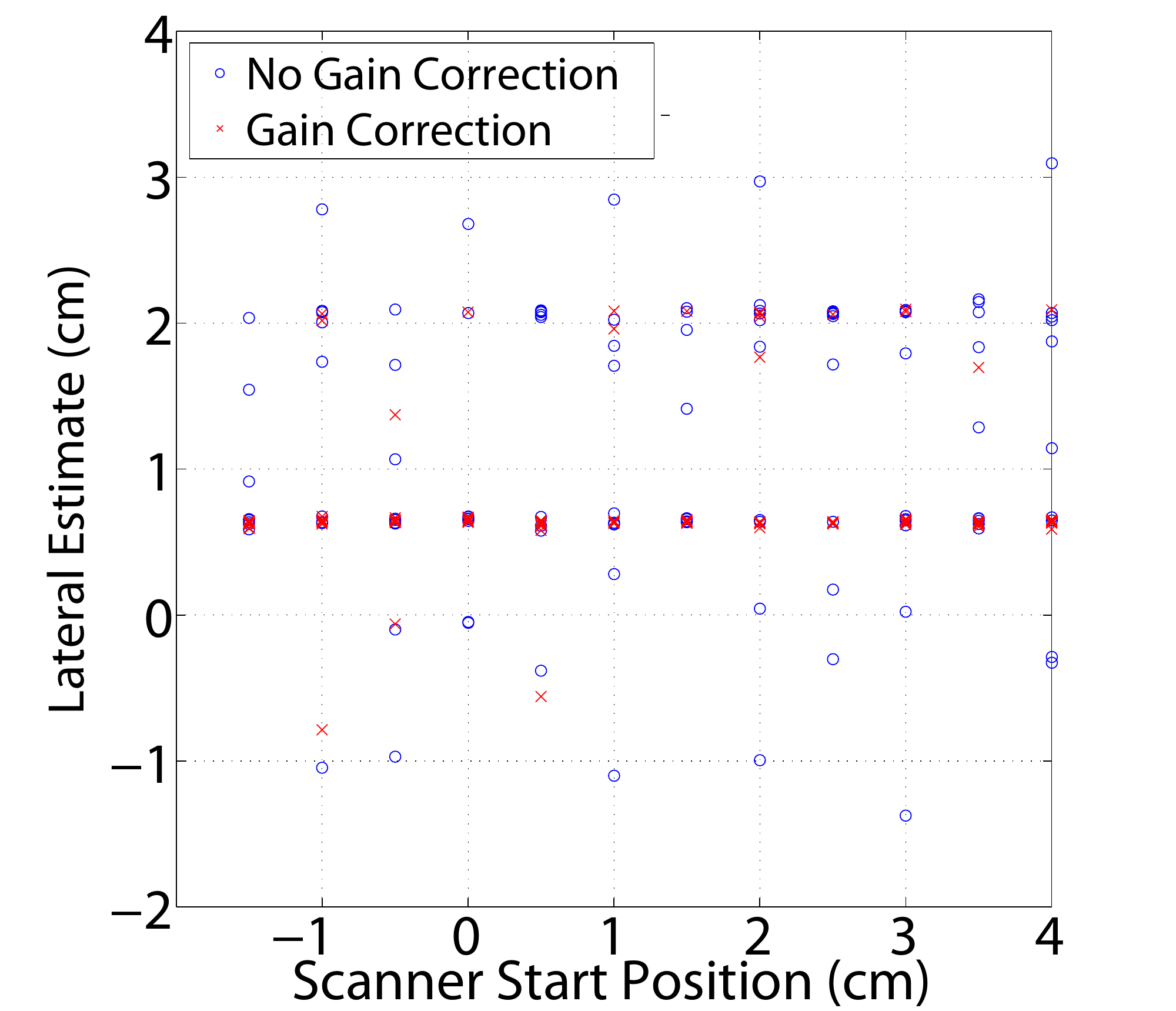}\\
(c) & (d)
\end{tabular}
\caption{The lateral estimates of a single point located at 0 mm from an optimal hopping window(a) and wide window(b) after eight lateral measurements, and of the weaker (greater distance from focus) of two point scatterers from an optimal hopping window (c) after six lateral measurements and a wide window (d) after seven lateral measurements.}
\label{fig:estimation_results}
\end{figure}

To ascertain the advantage gained from adaptive sampling, consider the number of simulated steps required for convergence for the optimal hopping and wide window cases with an object at a depth of 5 cm.  In the optimal hopping window case, the scanner could be started from as far away as 9 mm laterally from the scatterer.  The scanner was started at each 1 mm increment in this range, and 10 trials were performed at each starting location, constraining the estimate to the green region of Fig. \ref{fig:optimal_geometry}.  In the wide window case, the scanner was started at each 2 mm increment between -2 cm and +2 cm, and 10 trials were performed at each starting location.  Fig. \ref{fig:estimation_results}(a) and (b) show the lateral estimates for all trials at all starting locations after 8 lateral measurements for the optimal hopping and wide window cases, respectively. The variation for different runs at the same starting position were caused by the optimizer which provided slightly different solutions in each iteration of each run, from which statistical inferences may be made. From the scatter plots it is clear that within 8 lateral measurements the estimate was confined to less than a $\pm$1 mm lateral resolution for 84\% of the runs for the wide window case and 87\% of the runs for the optimal hopping window.  Considering all trials, adaptive sampling produces an approximate 30\% reduction in sampling for the optimal hopping window, given that the leftmost corner of the green parallelogram in Fig. \ref{fig:optimal_geometry} is 9 mm to the left of the beam center and the rightmost corner is 12 mm to the right of the beam center for an approximately 21 mm field-of-view.  Because the wide window is larger, adaptive sampling produces a much greater 60\% reduction in sampling. We note that the gain correction of Eq. \eqref{eq:gain_correction} had little effect on the single scatterer results. With no interfering scatterer present, accurate phase information was sufficient to locate a single scatterer.  

The fact that the $\pm$1 mm resolution could be achieved is of great significance.  The beam waist for the mean frequency sampled was $2.6$ mm. This nearly factor of three enhancement is a benefit of a recent insight in compressed sensing to move ``off of the grid"\cite{Recht_cs_off_grid}.  The non-linear model specified in Eq. \eqref{eq:no_ikx_obj} estimated the location of the scatterer in continuous space, for which the resolution is determined by the quality and number of measurements taken, not a sampling grid determined by the linear unbiased estimation bounds of the space-bandwidth product\cite{Brady}.  

Simulated experiments using real data were next run for both cases to locate two point scatterers: scatterer 1 at $x'_1=0.6$, $z'_1=6.9$ cm and scatterer 2 at $x'_2=2.1$, $z'_2=3.8$ cm. Because of its greater depth, the optimal hopping window would see scatterer 1 first, and according to Fig. \ref{fig:optimal_geometry} the acceptable starting locations of the scanner were from 12 mm to 3 mm to the left of  scatterer 1 ($i.e.$ -6 mm to +3 mm in the coordinates of the dataset). Again, the scanner was started at 1 mm increments within this range and 10 trials were performed at each starting location. For comparison, the wide window case searched a range 5.5 cm wide spanning the range -1.5 to 4 cm in the coordinates of the dataset.  Fig. \ref{fig:estimation_results} (c) and (d) show the lateral estimates for scatterer 1, which is deeper and scatters more weakly, for all trials at all starting locations after 6 or 7 lateral measurements for the optimal hopping or wide window cases, respectively. The optimal hopping window confines 93\% of the estimates for the weaker scatterer (scatterer 1) and 83\% of the estimates for the stronger scatterer (scatterer 2) to within 1 mm laterally after 6 measurements.  This represents a 74\% reduction in lateral sampling for the optimal hopping window. The wide window confined 84\% of the estimates for scatterer 1 and all of the estimates for scatterer 2 to within 1 mm laterally after 7 measurements.  This slightly larger window required one more measurement and therefore maintained the same lateral sampling reduction of 74\%.  

The gain correction in Eq. \eqref{eq:gain_correction} was needed for good estimation accuracy of the two point scatterers.  If the amplitude of one scatterer could not be accurately removed, some of the phase from it would be left in the data while estimating the location of the other scatterer, thereby producing inaccurate results.  While the locations of the scatterers are estimated concurrently, the data amplitude is dominated by the stronger scatterer, so estimating its location is more important for minimizing the objective function.  In the wide window case, the locations of the two scatterers were unconstrained within the volume.  Without the gain correction, the weaker scatterer was often estimated to be at the location of the stronger scatterer (2.1 cm laterally) due to this latent phase problem.  This problem was greatly reduced by gain correcting the data to match the model $M_0$ more accurately.  The optimal hopping window case did not suffer as greatly from latent phase since the weaker scatterer was detected first, and its lateral location therefore strongly constrained by the geometry of Fig. \ref{fig:optimal_geometry}.  Nevertheless, its gain corrected estimates exhibited a lower variance than the uncorrected estimates.

Since these simulated scans assumed that the number of scatterers and estimation parameters was known exactly, a final simulated experiment was performed to estimate the robustness of these cases for an over-specified or under-specified number of scatterers.  In both cases the algorithm succeeded in spite of the mis-specification.  As an example of the over-specified case, consider the previous single scatterer experiments but allow the algorithm to scan adaptively through the data mistakenly assuming there are two scatterers present. Ten trials were performed for each starting location, beginning with the scanner 2 cm to the left of the scatterer.  Each time, after 8 lateral measurements the scanner was able to find the true location of the single scatterer.  Although the algorithm tried to find a second scatterer, in 80\% of the trials it estimated this nonexistent scatterer to be in the same location as the first, albeit weaker by an average of 6 dB.  As an example of the under-specified case, consider the previous two scatterer experiment but assume there is only one scatterer.  Ten adaptive scanning trials were performed for each starting location, each time starting the scanner 2 cm to the left of the weaker scatterer.  After 8 measurements the weaker scatterer was always located to within 1.3 mm laterally.  The slight loss in resolution can be attributed to the interference from the other scatterer which is not correctly taken into account. Overall, the adaptive algorithm can be considered robust, even when the number of points is mis-specified.

\subsection{Fully Adaptive Scans}
A final verification of the method was performed by adaptively driving the scanner itself during acquisition.  The goals were the same as in the simulated adaptation case: to ascertain how many lateral scan measurements are required to locate a single scatterer and to what accuracy its position may be found.  To generate these statistics, the scatterer was started at various locations within the fixed field.  In a scenario similar to the wide window case presented in Sec. \ref{sec:simulations}, thirty adaptive measurements were made for each starting position in a window laterally constrained to be within 2 cm of the starting location and to a depth between 0 - 9 cm from the focus.  A typical adaptive path is shown in Fig. \ref{fig:typical_run}(c).  Clearly the algorithm followed a similar decision pattern to the simulated adaptation runs and quickly converged on the lateral location of the scatterer.  The convergence of det($\AN$) also followed the same pattern as shown in Fig. \ref{fig:typical_run}(d), starting out large and becoming smaller as the estimate converges to zero in approximately seven measurements.  

\begin{figure}
 \centering
 \begin{tabular}{cc}
 \includegraphics[width=2.5in]{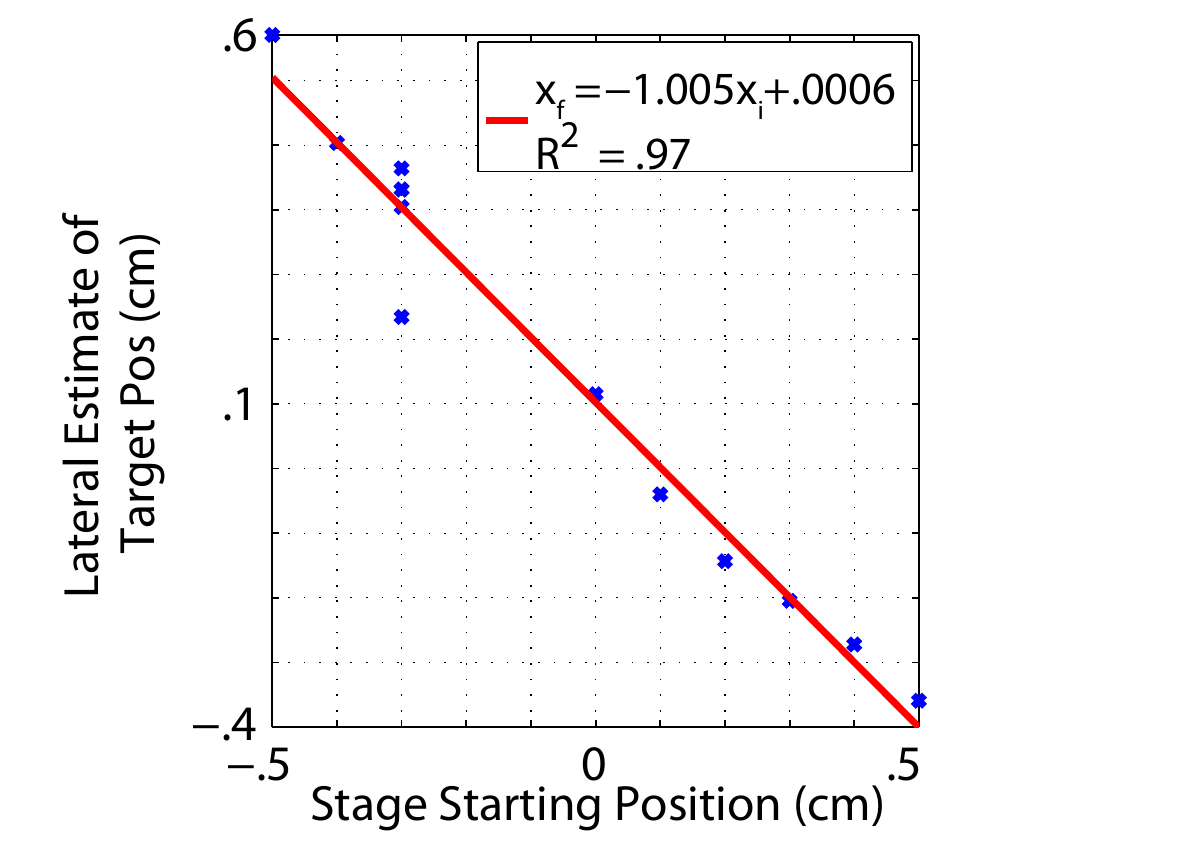} &
 \includegraphics[width=2.55in]{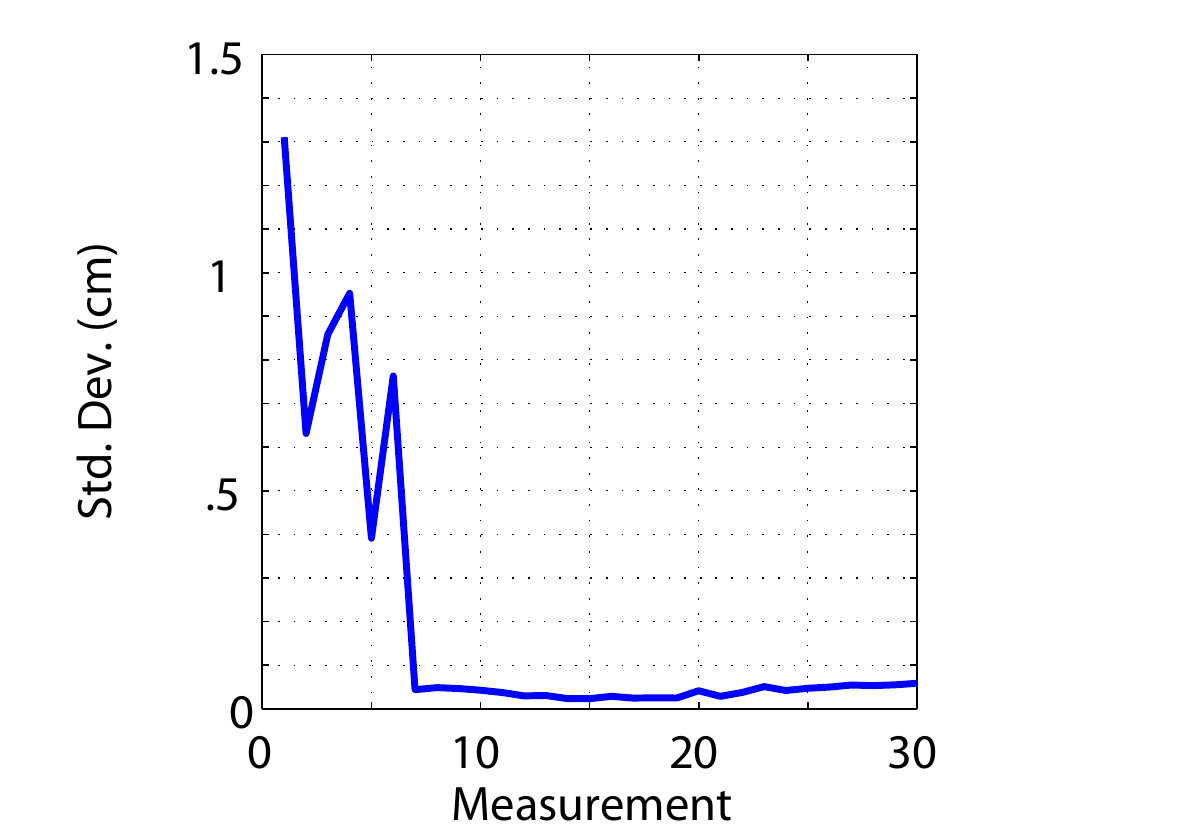}\\
 (a) & (b)
 \end{tabular}
 \caption{The regression line specifying the true location of the target (a) and the standard deviation from that line vs. iteration (b) for the fully adaptive experiment.}
 \label{fig:full_scan}
 \end{figure}

After all the runs, the set of final lateral estimates $x_f$ were plotted against the associated relative initial starting locations $x_i$, and a regression line was fit to ascertain how close the algorithm came to the ground truth $x_f = -x_i$.  (The coordinate system of the stage is flipped relative to the coordinate system of the estimation code, hence the negative slope.)  The regression line shown in Fig. \ref{fig:full_scan}(a) reveals how close the algorithm came to this ideal, especially given that the wire was offset a fraction of a millimeter from the stage center $x=0$.  Using fits like this for each iteration, the rate of convergence and accuracy of the algorithm could be estimated by measuring the standard deviation of the data from the regression line.  This standard deviation, shown in Fig. \ref{fig:full_scan}(b), shows that the algorithm converged after 7 lateral measurements with a $1\sigma$-error of $\sim\pm$ $0.5$mm.  Again, it is quite significant that a resolution of $<$ 1 mm was achieved using this ``off of the grid" nonlinear model, just as it was in the simulated experiments assuming the $2\sigma$-error as the resolution.  The number of measurements required for a given resolution were thereby dramatically reduced:  if the linear unbiased estimator were able to achieve this resolution, it would require approximately three times as many lateral measurements.

\section{Implications and Assumptions}
The results presented here were for specific cases, showing resolution enhancement and reduction in the number of lateral measurements for synthetic aperture sensing.  Although it is difficult to generalize these findings, rules of thumb from compressed sensing and the observations made here can at least provide some intuition.  Compressed sensing theory is largely based on RIP which states that all measurements should give equal information about the scene ($i.e.$ measure approximately the same amount of energy).  For this to be true, the measurements must be widely distributed across the scene.  In the optimal hopping window case this is indeed true.  The farther away the volume of interest is from the focus, the wider the green parallelogram of Fig. \ref{fig:optimal_geometry} will be.  This will allow for greater speedups in acquisition, but it comes at the cost of reduced SNRs.  In turn, the ability to super-resolve the points will be degraded since small variations in the phase will be less distinguishable.  Widening the window of estimation also allows for more compressed estimation of the object locations.  However, in this case there is a finite probability of observing no energy, so RIP is violated.  Our work shows that with good SNR this is acceptable to some extent; however, a system designer should be aware that expanding the window too far could lead to bad estimates.  Exactly how far is too far largely depends on SNR and the number of points needed to be estimated simultaneously.  Ascertaining these dependencies on the salient variables in this more generic problem will be addressed in a subsequent project.

It has been shown that incorrectly specifying the number of scatterers to be located does not prevent the algorithm from correctly locating all the scatterers in the field for the over-specified case or the specified number of scatterers in the under-specified case.  However, it would be preferable to select the number of scatterers to be located adaptively.  This is a well-known problem related to the relevance vector machine \cite{RVM} which has been related to compressed sensing and adaptive sensing by \cite{bayesian_cs}.  A disadvantage of this strategy is that the parameter space is as large as the inherent dimension of the estimation space, thus requiring inversions of large matrices in each step.  Furthermore, the grid must once again be fixed for adaptive selection of the scene sparsity, removing the ability to have the resolution of the estimate be determined by the quality of the measurement system.  Perhaps the resolution of the grid could first be determined by an adaptive measurement scheme, such as the one presented here, so that the grid can be accurately specified for the relevance vector machine strategy.

The remaining assumptions are that the beam's spatial distribution M(x,k) is Gaussian and may be demodulated by referencing the wavevector to the minimum frequency used ($k \rightarrow k - \kmin$).  The case of a non-Gaussian spatial profile is easily addressed by changing M(x,k) to reflect the beam's actual spatial profile, assuming this is known accurately.  In this case, the algorithm should still identify the correct locations of the scatterer(s), even if the strongest signals occur when the scatterer is not in the center of the beam.  By contrast, the demodulation correction was not made ab initio but in response to the quality of the convergence in the fits.  It is likely that different scenarios will require different versions of this demodulation, but the approach used here will work for most non-dispersive scatterers since their behaviors are similar at base band as at the operational frequencies.  The correction of the beam gain by Eq. \eqref{eq:gain_correction} was in response to the experimental data recognizing that phase may not be altered and appropriately compensating the amplitude to match the antenna performance.  Similar corrections may be required for other antenna structures.

Perhaps the most challenging assumptions involve the scatterers themselves; namely, that their scattering cross sections do not depend sensitively on angle, frequency, or polarization, and that the embedding medium scatters much more weakly by comparison.  By using vertically oriented, sub-wavelength diameter wires as scatterers in two dimensions (x and z) and performing one-dimensional lateral scans, the dependence of scattering on angle and frequency in the measurement plane is removed, as is the polarization dependence because the source polarization was aligned with the wires.  Generalizing our findings to the case of a three dimensional array of sparse, oriented, non-spherical scatterers, we see that all three effects may vary significantly as the aspect of the scatterer changes over the course of the two dimensional scan.  These effects can be minimized by ensuring that the wavelength is significantly larger than the size of the scatterers and that circular polarization is used instead of linear.  Such scatterers even more strongly demand a well conditioned Gaussian beam to simplify the analysis. Nevertheless, this work has shown that compressive sampling techniques are practical and may become increasingly indispensable for MMW and THz imaging applications, especially when used to locate or render scenes dominated by a few sparsely-arranged scatterers. 

\bibliographystyle{ieee}


\end{document}